\documentclass[twocolumn]{aastex631}

\usepackage{soul}
\usepackage{amsmath}
\usepackage{booktabs}

\newcommand{\ci}{[\ion{C}{1}]}
\newcommand{\cii}{[\ion{C}{2}]}

\newcommand{\msun}{\mbox{$\,\rm M_{\odot}$}}

\newcommand{\htwo}{\mbox{H$_2$}}

\newcommand{\mmol}{\mbox{$M_{\rm mol}$}}
\newcommand{\mstar}{\mbox{$M_{*}$}}

\newcommand{\lpci}{\mbox{$L'_{\rm CI}$}}
\newcommand{\alphaci}{$\alpha_{\rm [CI]}$}

\received{-}
\revised{-}
\accepted{-}

\submitjournal{ApJ}

\shorttitle{GRB hosts: deep \ci-nondetections}
\shortauthors{Nadolny et al.}

\begin{document}

\title{
Main Sequence to Starburst Transitioning Galaxies:\\
Gamma-ray Burst Hosts at $z\sim2$
}


\correspondingauthor{Jakub Nadolny}
\email{jakub.nadolny@amu.edu.pl}

\author[0000-0003-1440-9061]{Jakub Nadolny}
\author[0000-0001-9033-4140]{Micha{\l}~Jerzy Micha{\l}owski}
\affiliation{Astronomical Observatory Institute, Faculty of Physics, Adam Mickiewicz University, ul.~S{\l}oneczna 36, 60-286 Pozna{\'n}, Poland}

\author[0000-0002-8443-6631]{J. Ricardo Rizzo}
\affiliation{ISDEFE, Beatriz de Bobadilla 3, 28040 Madrid, Spain}
\affiliation{Centro de Astrobiolog\'ia (CAB), CSIC-INTA, Ctra.~M-108, km.~4, 28850
              Torrej\'on de Ardoz, Spain}

\author[0000-0001-8913-925X]{Agata Karska}
\affiliation{Institute of Astronomy, Faculty of Physics, Astronomy and Informatics, Nicolaus Copernicus University, Grudziadzka 5, 87-100 Toru{\'n}, Poland}
\affiliation{Max-Planck-Institut für Radioastronomie, Auf dem H\"ugel 69, 53121, Bonn, Germany}

\author[0000-0002-3947-1518]{Jesper Rasmussen}
 \affiliation{Technical University of Denmark, Department of Physics, Fysikvej 309, DK-2800 Lyngby, Denmark}
 
\author[0000-0003-1546-6615]{Jesper Sollerman}
\affiliation{ Department of Astronomy, The Oskar Klein Centre, Stockholm University AlbaNova, SE-106 91 Stockholm, Sweden}
 
\author[0000-0002-4571-2306]{Jens Hjorth}
\affiliation{ DARK, Niels Bohr Institute, University of Copenhagen, Jagtvej 128, 2200 Copenhagen, Denmark}
 
\author[0000-0002-8860-6538]{Andrea Rossi}
 \affiliation{ INAF - Osservatorio di Astrofisica e Scienza dello Spazio, via Piero Gobetti 93/3, 40129 Bologna, Italy}
 

\author[0000-0002-3148-1359]{Mart\'in Solar}
\author[0000-0002-3007-1868]{Rados{\l}aw Wr\'oblewski}
\author[0000-0001-8723-3533]{Aleksandra Le\'sniewska}
\affiliation{Astronomical Observatory Institute, Faculty of Physics, Adam Mickiewicz University, ul.~S{\l}oneczna 36, 60-286 Pozna{\'n}, Poland}



\begin{abstract}

Star-forming galaxies populate a main sequence (MS), a well-defined relation between stellar mass (\mstar) and star-formation rate (SFR). Starburst (SB) galaxies lie significantly above the relation whereas quenched galaxies lie below the sequence. In order to study the evolution of galaxies on the SFR-{\mstar} plane and its connection to the gas content, we use the fact that recent episodes of star formation can be pinpointed by the existence of gamma-ray bursts (GRBs). Here we present sensitive \ci-nondetections of $z\sim\,2$ ultra luminous infrared (ULIRG) GRB host galaxies. We find that our GRB hosts have similar molecular masses to those of other ULIRGs. However, unlike other ULIRGs, the GRB hosts are located at the MS or only a factor of a few above it. Hence, our GRB hosts are caught in the transition toward the SB phase. This is further supported by the estimated depletion times, which are similar to those of other transitioning galaxies. The GRB hosts are \ci-dark galaxies, defined as having a \ci/CO temperature brightness ratio of $<$0.1. Such a low \ci/CO ratio has been found in high-density environments ($n_{\rm H}>10^4\,{\rm cm}^{-3}$) where CO is shielded from photodissociation, leading to under-abundances of \ci. This is consistent with the merger process that is indeed suggested for our GRB hosts by their morphologies.

\end{abstract}


\section{Introduction} 
\label{sec:intro}
Star formation occurs in molecular gas clouds (\citealt{Wong2002ApJ...569..157W,Gao2004ApJS..152...63G,Bigiel2008AJ....136.2846B}; but see \citealt{Glover2012MNRAS.421....9G,Krumholz2012ApJ...759....9K,Michalowski2015A&A...582A..78M}). The molecular gas fraction and its availability for star formation are key ingredients that shape the evolution of galaxies \citep[see review by][]{Saintonge_annurev-astro-021022-043545} and determine the place where a galaxy is found in the star-formation rate (SFR)--stellar mass ($\mstar$) plane.  Normal star-forming galaxies (SFGs) form a well-defined `main sequence' (MS) on this plane
with a scatter of about 0.2 dex \citep{Brinchmann2004MNRAS.351.1151B,Noeske2007ApJ...660L..43N,Speagle2014ApJS..214...15S}.

The so-called starburst galaxies are found above the MS with extremely high SFRs for a given stellar mass \citep{Combes2011A&A...528A.124C,Rodighiero2011ApJ...739L..40R,Larson2016ApJ...825..128L}. It is not clear whether an increase in star formation efficiency \citep[SFE;][]{Cheng2018MNRAS.475..248C,Hogan2022MNRAS.512.2371H} or gas mass fraction \citep{Lee2017MNRAS.471.2124L,Valentino2020A&A...641A.155V} drives the departure from the MS. As pointed out by \cite{Gao2004ApJ...606..271G}, the global SFR depends mainly on the amount of dense molecular gas, which can be traced for example by a hydrogen cyanide (HCN) line. This dependence remains nearly the same (with a slope of about 1) for normal and starburst galaxies, including ultra-luminous galaxies (ULIRGs). The HCN observations however are limited to the local universe due to the weakness of this line. Thus, until deep HCN (or other dense gas tracers) observations are available, we need to rely on other approximations.

Major and minor mergers have been invoked as possible causes for triggering the starburst behavior \citep{Combes2011A&A...528A.124C,Rodighiero2011ApJ...739L..40R,Larson2016ApJ...825..128L,Saintonge_annurev-astro-021022-043545}. Eventually, some galaxies will terminate their star formation. These galaxies tend to have red colors, compact and spheroidal morphologies \citep{Schawinski2014MNRAS.440..889S,Nadolny2021A&A...647A..89N}, with relatively low gas and dust content.

Most of the information on molecular gas in galaxies comes from observations of the carbon monoxide (CO) lines \citep{Bolatto2013ARA&A..51..207B,Carilli2013ARA&A..51..105C}. The notion that the neutral carbon line (\ci) traces the bulk of the molecular gas mass has been investigated for more than four decades now \citep{Phillips1981ApJ...251..533P,Papadopoulos2004MNRAS.351..147P,Jiao2017ApJ...840L..18J,Valentino2018ApJ...869...27V}. 
On the other hand, in UV-intense and metal-poor environments, the use of \ci\ as a molecular gas tracer is limited due to the increased ionisation of carbon. The limited usefulness of \ci\ has also been shown in dense conditions, i.e in the collisional fronts of mergers \citep{Michiyama2021ApJS..257...28M}. \cite{Bisbas2017ApJ...839...90B} showed however, that even if \ci\ is limited in such cases, it is still a more reliable tracer of global molecular mass than CO. {This is due to the high sensitivity of the [C/CO] abundance to even small changes in the cosmic ray ionization rate, especially when the average gas densities are low ($\le 10^3 {\rm cm}^{-3}$). It is in such densities that the bulk of the H$_2$ reservoirs in galaxies is often found.}



Recent episodes of star formation in galaxies can be pinpointed by the existence of gamma-ray bursts (GRBs) which are explosions of short-lived massive stars \citep{Hjorth2003Natur.423..847H,Stanek2003ApJ...591L..17S}. 
In this paper we take advantage of this feature to study the evolution of galaxies on the SFR-\mstar\ plane and its connection to the gas content by analyzing high-sensitivity observations of \ci\ line emission towards selected GRB hosts.  In particular, we want to shed light on the cause of weak \ci\ lines of galaxies, i.e.~if they are due to gas properties (e.g., density) or due to true low molecular gas content.

This paper is organized as follows. In Section \ref{sec:data} we describe the sample selection, observations, and reduction process. We also define the comparison sample from the literature and describe the methods used to derive fluxes, luminosities, and molecular masses. In Section \ref{sec:results} we describe the results of our analysis. Section \ref{sec:grb_hosts} discusses our interpretation of the observables together with alternative scenarios. Finally, in Section \ref{sec:discussion_and_conclusions} we present the conclusions of our work. 
Throughout this paper we use a cosmological model with $H_0$ = 70 km s$^{-1}$Mpc$^{-1}$, $\Omega_\Lambda = 0.7$, and $\Omega_m = 0.3$.

\section{Data} \label{sec:data}
\subsection{Sample Selection}
The GRB host sample observed in \ci($^3P_1 - ^3P_0$) (hereafter \ci) was selected  based on the availability of infrared or radio detections \citep{Hunt2014A&A...565A.112H,Perley2015ApJ...801..102P,Michalowski2015A&A...582A..78M}, allowing precise estimates of SFRs. We selected hosts with spectroscopic redshifts so that their {\ci} lines were expected to be observed away from atmospheric water lines.
This resulted in seven potential targets (the hosts of GRB\, 051006, 051022, 060814, 061121, 080207 100316D, 111005A). Depending on their declinations, these sources were observed with the Atacama Pathfinder Experiment (APEX) or the IRAM 30m radio telescope. The hosts of GRB\,111005A, 051006, 051022 were not observed, because they are at the lower redshift range proposed for a given telescope, resulting in a high observing frequency at which the weather requirements were challenging.

Low-redshift targets (GRB\,061121 and 100316D) were only suitable for the APEX telescope because for them the observing frequency is high and requires very stable weather conditions and also a very low amount of precipitable water vapor. These stringent conditions are often attained at Chajnantor (the APEX site), but not at Pico Veleta (the IRAM 30m telescope site).

\subsection{Observations and Data Reduction}

\begin{figure*}[ht]
\begin{center}
\includegraphics[width=0.9\columnwidth,clip]{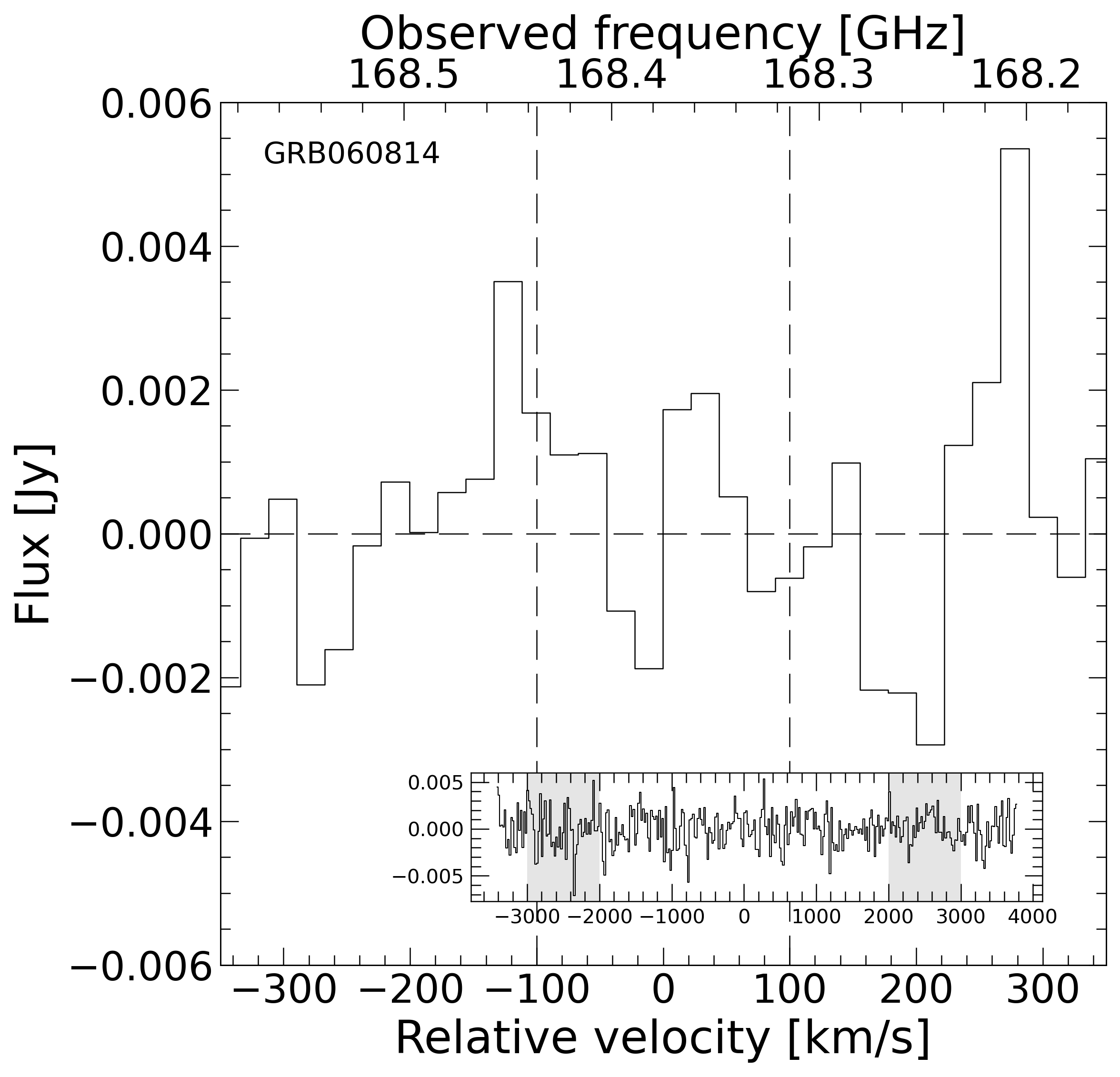}
\includegraphics[width=0.91\columnwidth,clip]{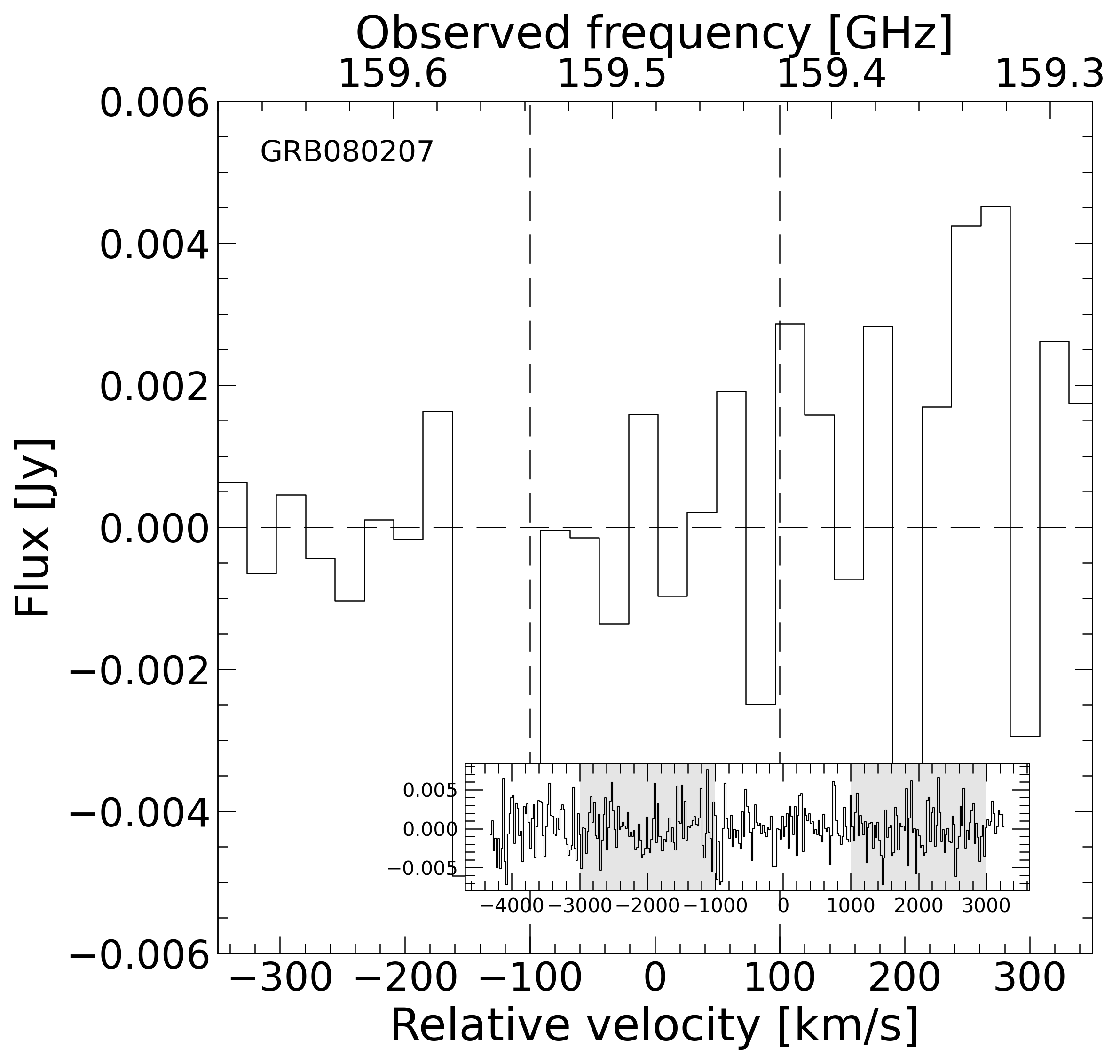}
\end{center}
\caption{The \ci\ spectra of GRB\,060814 and 080207. Vertical dotted lines indicate the velocity range used for the flux integration (from $-100$ to $100$ km s$^{-1}$). The upper x-axis labels the observed frequencies. Inset panels show the whole obtained spectra, while shaded areas indicate ranges used to estimate the noise. \label{fig:ci_spectra}}
\end{figure*}

We observed the hosts of GRB\,060814 and 080207 with the IRAM 30m  telescope (proposal 172-16; PI: M.J.M.), equipped with the Eight MIxer Receiver\footnote{ \url{www.iram.es/IRAMES/mainWiki/EmirforAstronomers}} \citep{Carter2012A&A...538A..89C}. 
We implemented the wobbler-switching mode and the Fourier Transform Spectrometers 200 (FTS-200) providing 195\,kHz spectral resolution  and 16\,GHz bandwidth in each linear polarisation. 
The observations for both targets were executed between 2017-Feb-01 and 2017-May-22 and lasted in total 13.1\,hr on-source for GRB\,060814 and 17.8\,hr for GRB\,080207.
The observations were divided into 6 min scans, each consisting of 12 subscans 30\,s long. Pointing was checked and corrected every 1--2 hr. 
Each spectrum was calibrated, and corrected for baseline shape. 
The spectra were aligned in frequency and noise-weight averaged. Some well-known platforming, due to the fact that the instantaneous bandwidth of 4 GHz is sampled by three different FTS units, was corrected off-line by a dedicated procedure within the Continuum and Line Analysis Single Dish Software ({\sc Class}). In all cases, the {\ci} line is far away from the step of the platforming. 

The hosts of GRB\,061121 and 100316D were observed with the APEX telescope (\citealt{Gusten2006A&A...454L..13G}; proposals 098.F-9300 and 098.D-0243; PI: M.J.M.), equipped with the Swedish Heterodyne Facility Instrument (SHeFI; \citealt{Vassilev2008A&A...490.1157V,Belitsky2006SPIE.6275E..0GB}).
However the upper limits were not sufficiently constraining to allow robust conclusions about the molecular content of these sources, hence we do not report the results for these galaxies. 

All data were reduced and analyzed using the {\sc Class} package within the Grenoble Image and Line Data Analysis Software: {\sc Gildas}\footnote{\url{www.iram.fr/IRAMFR/GILDAS}} \citep{Pety2005sf2a.conf..721P}.

The obtained \ci\ spectra of our two GRB (060814 and 080207) hosts are shown in Figure \ref{fig:ci_spectra}.

\subsection{Comparision Sample}
In order to place our two GRB hosts into a general perspective we compiled a sample of galaxies spanning several orders of magnitude in stellar mass, SFR and gas mass. In particular, this compilation contains normal SFG \citep{Bourne2019MNRAS.482.3135B,Dunne2021MNRAS.501.2573D,Valentino2020ApJ...890...24V}, (Ultra) Luminous IR Galaxies [(U)LIRGs] \citep{Valentino2020ApJ...890...24V,Michiyama2021ApJS..257...28M,Lu2017ApJS..230....1L}, intermediate-$z$ isolated LIRGs \citep{Lee2017MNRAS.471.2124L} and merging starburst ULIRGs \citep{Combes2011A&A...528A.124C}, high-$z$ starbursts \citep{Shi_KS_outliers_2018}, and transitioning isolated and merging galaxies \citep{Cheng2018MNRAS.475..248C,Hogan2022MNRAS.512.2371H}. For all sources with \ci\ data, \mmol\ has here been estimated in a consistent manner (see Sec. \ref{sec:fluxes}).  
We used the total far-infrared luminosity to estimate the SFR \citep{Kennicutt1998ARA&A..36..189K}, as \[SFR=L_{\rm IR}[L_\odot]/(9.86 \times 10^9) \rm{M}_\odot \rm{yr}^{-1} \](converted to the \citealt{Chabrier2003ApJ...586L.133C} IMF), and the fundamental plane \citep{Maritza_FP_2010A&A...521L..53L} to estimate the oxygen metallicity \citep{Valentino2018ApJ...869...27V,Bourne2019MNRAS.482.3135B,Dunne2021MNRAS.501.2573D}, for galaxies without these estimates in the literature. The latter is needed to estimate the metallicity-dependent conversion factor \alphaci\ introduced below.


\subsection{Flux, Luminosity and Mass Measurements}
\label{sec:fluxes} 
Our deep observations of \ci\ in GRB hosts show a lack of significant emission. We estimated $2\sigma$ upper limits of the line flux, luminosity, and molecular gas masses by integration within a velocity range of -100 to 100 km s$^{-1}$ around the expected velocity of the \ci\ line at the redshift of each source. For our sample, as well as for the data from other works, we estimated the \ci\ luminosity \lpci\ using the same prescription -- equation 3 in \citet{Solomon1997ApJ...478..144S}. 

We employed two methods to estimate the molecular gas mass from the luminosity of the {\ci} line. The first method is based on a theoretical analysis of the \ci\ emission from the ISM assuming local thermal equilibrium given by \cite{Papadopoulos2004MNRAS.351..147P}. Using their Equation 11, evaluating all the constants we derived an expression to estimate the molecular gas mass from the \ci(1-0) line flux as:

\begin{equation}
\begin{split}
    \frac{M_{\rm mol}}{\rm M_{\odot}} = 1.3747 \times 10^{-9} \frac{D_{\rm L}^2}{1 + z} \frac{I_{\rm CI}}{(X_{\rm CI} A_{\rm 10} Q_{10})},
\end{split}
\end{equation}
where the abundance ratio is $X_{\rm CI} = 3\times10^{-5}$  \citep{Valentino2018ApJ...869...27V,Jiao2017ApJ...840L..18J}, Einstein coefficient for this transition is $A_{10} = 7.93 \times 10^{-8}$s$^{-1}$, $Q_{10}$ is given by Equation A15 from \cite{Papadopoulos2004MNRAS.351..147P} with assumed $T_{\rm kin} = 40\rm{\ K}$, $D_{\rm L}$ is the luminosity distance given in Mpc, and $I_{\rm CI}$ is the velocity integrated \ci\ line flux in units of Jy\,km\,s$^{-1}$. We refer to this method as P04 (see \citealt{Weiss2003A&A...409L..41W} for a similar method).
The second method is taken from \citet{Heintz2020ApJ...889L...7H}, and is based on the conversion factor \alphaci\, between \mmol\ and \lpci, estimated from observations of the \ci\ absorption line in spectra of GRB afterglows and QSOs. In this case \alphaci\ is metallicity dependent. 
We refer to this method as H20. Note that in both methods a factor $\mu=1.36$ that corrects for the contribution from helium and heavier elements is included.

\section{Results} \label{sec:results}
\begin{table*}[ht]
\centering
\caption{Results from our {\ci} observations. The \ci\ fluxes are given with their 2$\sigma$ errors, while luminosities and molecular masses are 2$\sigma$ upper limits. The errors were obtained by randomly perturbing 1000 times the fluxes of each channel in the spectra within their errors and assessing the 95\% confidence interval of the obtained integrated fluxes. The measurements of \mmol\ for all the methods include the $\mu = 1.36$ factor accounting for the contribution from helium and heavier elements. The last column (M18) reports CO-based molecular mass from \citet{Michalowski2018_CO_GRB}.\label{tab:measurments}}
\medskip
\begin{tabular}{cccrrr}
\hline
GRB  & $I_{\rm[CI]}$     & $\log($\lpci$)$   &   \multicolumn{3}{c}{$\log(M_{\rm mol}/M_{\odot})$}\\\cmidrule{4-6}
    & (Jy km s$^{-1}$)    &  (K km s$^{-1}$ pc$^{-2}$)   & P04$^{(a)}$ & H20$^{(a)}$ & M18\\
\hline
060814 & -0.064 $\pm0.113$ & 9.178 & 9.852 & 10.859 &  10.92 \\
080207 & 0.016 $\pm0.139$ & 9.504 & 10.17 & 10.777 &  11.3 \\
\hline
\end{tabular}
\tablecomments{$^{(a)}$Molecular gas masses from this work using methods P04 and H20.}
\end{table*}


\begin{table}[ht]
\begin{center}
	\caption{GRB host properties from the literature.
	\label{tab:sources_info}}
\begin{tabular}{cccc}
GRB  & $z$  & SFR (M$_\odot$/yr) & $\log$(M$_*$/M$_\odot$)\\
\hline
060814 & 1.92 (H12) & 256.0 (P15) & 10.2 (P15) \\
080207 & 2.09 (H12) & 170.0 (H12) & 11.17 (H14)\\
\hline
\end{tabular}
\tablecomments{References in parenthesis are H12: \cite{Hjorth2012ApJ...756..187H}; H14: \cite{Hunt2014A&A...565A.112H}; P15: \cite{Perley2015ApJ...801..102P}}
\end{center}
\end{table}

All the measurements obtained for our GRB hosts are given in Table \ref{tab:measurments}. The upper limits are 2$\sigma$. The SFRs, stellar masses, and spectroscopic redshifts from the literature are given in Table \ref{tab:sources_info}.

\subsection{[C\ I] Luminosity}

\begin{figure}
\includegraphics[width=0.95\columnwidth,clip]{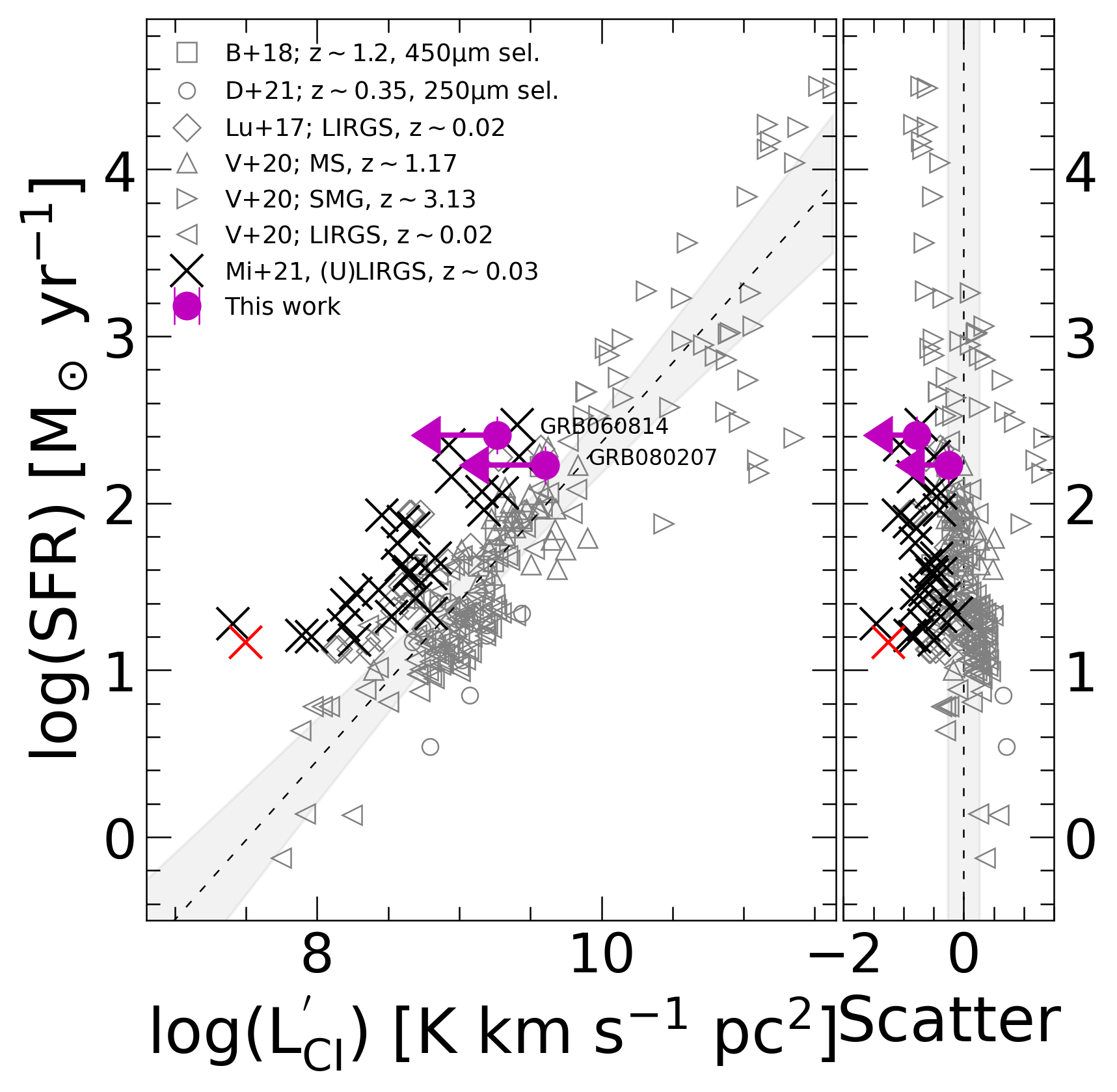} 
\caption{Star-formation rate as a function of [CI] emission line luminosity \lpci. The dashed line and shaded region mark our linear fit (Eq. \ref{eq:sfr_lpci}) and its $2\sigma$ scatter, respectively, based on data (gray open markers) from the literature  \citep{Bourne2019MNRAS.482.3135B,Dunne2021MNRAS.501.2573D,Lu2017ApJS..230....1L,Valentino2020ApJ...890...24V}. The right panel shows the residuals relative to the best-fit relation. The red X symbol shows the \ci\ upper limit of \ci-dark galaxy found in \cite{Michiyama2021ApJS..257...28M}. See main text. \label{fig:lpci_sfr}}
\end{figure}

In Figure \ref{fig:lpci_sfr} we show the SFRs as a function of \ci\ line luminosity \lpci\ $[{\rm K\,km\,s}^{-1} {\rm pc}^{-2}]$ for our sample and for data from the literature for which the \ci\ fluxes were available \citep{Bourne2019MNRAS.482.3135B,Lu2017ApJS..230....1L,Valentino2020ApJ...890...24V,Dunne2021MNRAS.501.2573D,Michiyama2021ApJS..257...28M}. Using only data for SFG from the literature (gray empty markers), we find the best-fit relation between \lpci\ and SFR as:
\begin{equation}
\label{eq:sfr_lpci}
\log(\mbox{SFR})[{\rm M_\odot}/{\rm yr}] = 0.952 \times \log({\rm \lpci}) -7.165
\end{equation}
which holds over a wide redshift range ($0\,<\,z\,<\,3$) with $2\sigma$ scatter of 0.27 dex. As shown in Figure \ref{fig:lpci_sfr}, the hosts of GRBs 060814 and 080207 have lower \lpci\ (by $>$0.8 and $>$0.3 dex, respectively) than expected for their SFRs based on our best-fit. 
It has been shown for GRB hosts, in general, that, we can rule out the possible contamination of the emission by active galactic nuclei at the wavelengths used to estimate SFRs \citep{Perley2015ApJ...801..102P}, so we consider these SFRs to be robust. 

In Figure \ref{fig:lpci_sfr} we also show the local (U)LIRG merging galaxies \citep{Michiyama2021ApJS..257...28M}, which together with our low-\lpci\ GRB hosts are found above the SFG population in terms of their SFR. This is also clearly visible on the right-hand panel in the same Figure where the scatter around the best-fit relation is shown. In particular, the red X symbol shows NGC 7679, the \ci-dark AGN host found by \cite{Michiyama2021ApJS..257...28M}. The \ci/CO ratio of 0.07 of this galaxy is below those of other ULIRGs \citep{Jiao2017ApJ...840L..18J}. Using the CO data from \cite{Michalowski2018_CO_GRB} we estimated the \ci/CO ratio of our GRB hosts to be of the order of $\leq$0.1. This is lower than normal ULIRGs but not as low as \ci-dark galaxies like NGC 7679 or NGC 6052 studied in \cite{Michiyama2020ApJ...897L..19M,Michiyama2021ApJS..257...28M}. On average (U)LIRGs have 0.6 dex lower \lpci\, while the \ci-dark NGC 7679 has 1.2 dex lower \lpci\ than the expectations from the \lpci-SFR fit.

\subsection{Molecular Gas Mass}
\begin{figure*}
\begin{center}
\includegraphics[width=.8\textwidth,clip]{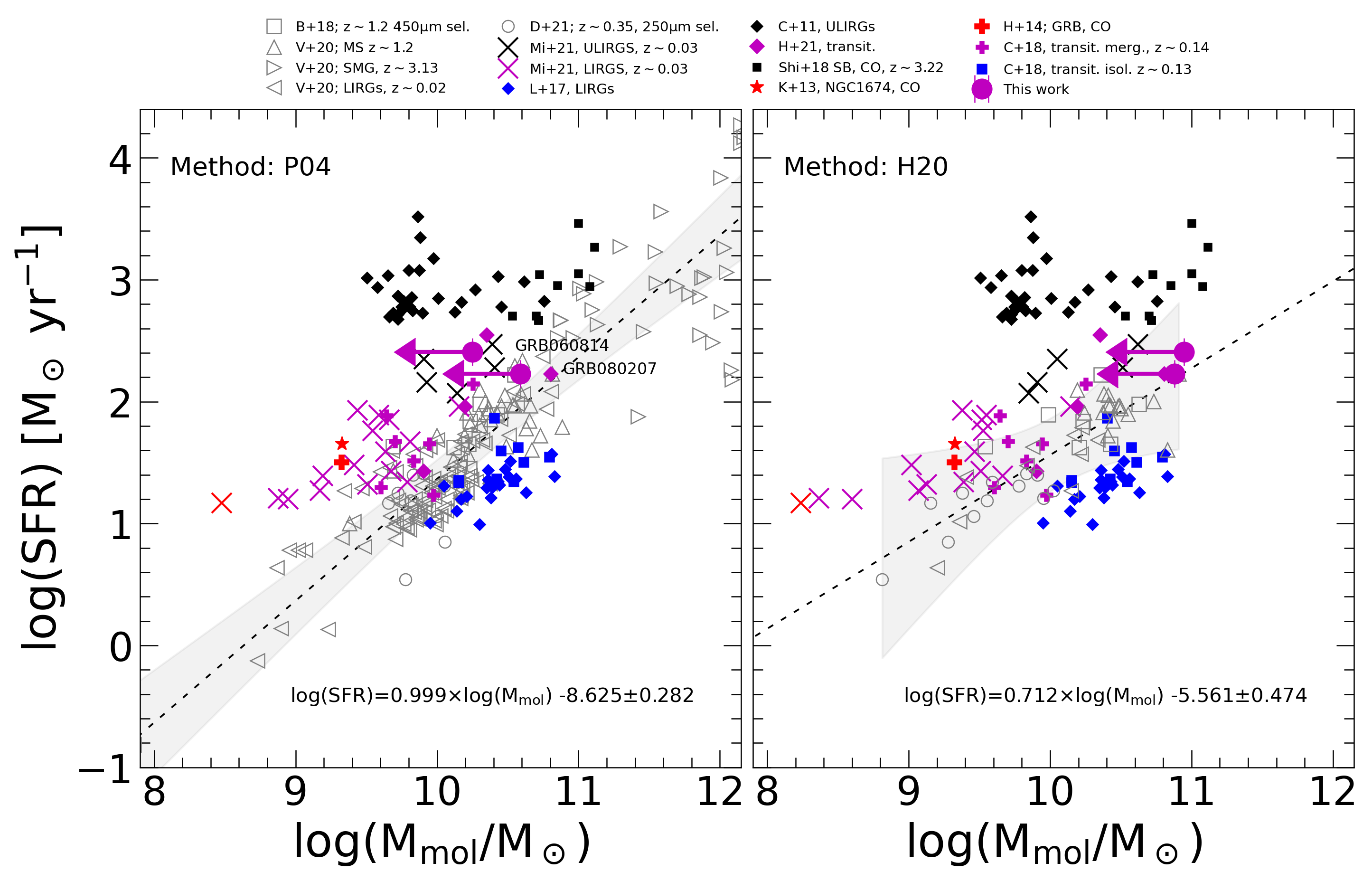}
\end{center}
\caption{Star formation rate as a function of molecular gas mass, \mmol, for the two estimation methods used (labeled in each panel). Magenta arrows show the upper-limit estimation of \mmol\ of GRB hosts from this work. Gray open markers represent the data from the literature used in the fitting \citep{Bourne2019MNRAS.482.3135B,Valentino2020ApJ...890...24V,Dunne2021MNRAS.501.2573D}. Black filled markers represent intermediate- and high-$z$ starbursts ULIRGs \citep{Combes2011A&A...528A.124C,Shi_KS_outliers_2018}; magenta X, diamond, cross symbols represent merging (U)LIRGs \citep{Michiyama2021ApJS..257...28M,Shangguan2019ApJ...870..104S}, transitioning intermediate-$z$ galaxies \citep{Hogan2022MNRAS.512.2371H}, transitioning local merging systems \citep{Cheng2018MNRAS.475..248C}, respectively; blue diamonds and squares represent intermediate-$z$ isolated LIRGs \citep{Lee2017MNRAS.471.2124L}, and local isolated transitioning galaxies \citep{Cheng2018MNRAS.475..248C}; red star and cross reprsenent local LIRG NGC 1674 \citep{Konig_NGC1764_MolMass2013} and the molecule-deficient GRB\,051022 host galaxy \citep{Hatsukade2014Natur.510..247H}. For these galaxies, we use the same \mmol\ in both panels (mainly due to a lack of \lpci\ estimates), and these are only for comparison. The results of the fitting are given in each panel.\label{fig:Mmol_sfr}}
\end{figure*}


 In Figure \ref{fig:Mmol_sfr} we show the estimated molecular gas masses of our GRB hosts based on \lpci\ using our two different methods (see section \ref{sec:fluxes} for details). We include the same data as in Figure \ref{fig:lpci_sfr},\footnote{Not all the galaxies from Fig.~\ref{fig:lpci_sfr} had metallicity estimates available, nor was it possible to estimate these using the fundamental metallicity relation \citep{Maritza_FP_2010A&A...521L..53L}. These galaxies could not be included in the method where the conversion factor is metallicity-dependent, i.e.~for Method H20.} together with local LIRG NGC 1674 \citep{Konig_NGC1764_MolMass2013}, intermediate-$z$ isolated LIRGs \citep{Lee2017MNRAS.471.2124L}, intermediate-$z$ merging starburst ULIRGs \citep{Combes2011A&A...528A.124C}, and high-$z$ starburst ULIRGs \citep{Shi_KS_outliers_2018}, and low- and  intermediate-$z$ transitioning isolated and merging galaxies \citep{Cheng2018MNRAS.475..248C,Hogan2022MNRAS.512.2371H} with the molecular mass estimated from CO emission (i.e.~these have the same \mmol\ in both panels, and are shown for comparison only). Using the SFG sample from the literature (gray empty markers) we find the best-fit relation between SFR and \ci-based \mmol\ for each of the methods. The results of the fitting are given in each panel. The \mmol\ for the GRB\,060814 and 080207 host galaxies are found to be lower by $>$0.8 and $>$0.2 dex from our best-fit for the P04 method. At least in the case of the host galaxy of GRB\,060814 the offset is in agreement with the distance from the best-fit estimated for (U)LIRGs ($\sim$0.8 dex), high-$z$ starburst ($\sim$0.7), NGC 1674 ($\sim$0.9 dex), and for the transitioning SFG mergers ($\sim$0.4 dex). 
 
 The range of the molecular gas mass in isolated SFG (blue markers), merging transitioning systems (magenta), and in starburst (black) is similar (between $10^{9.5}$ and $10^{11}$\msun), but their SFRs vary significantly from tens to thousands solar masses per year, which translates to different depletion times (or SFEs; see Sec. \ref{sec:mainsequence_deptime}). Considering the second method used (H20), which is metallicity-dependent, we can see that the majority of the ULIRGs, including our GRB hosts, have \mmol\ above, but within the scatter of, the best-fit relation. 

\subsection{Main Sequence and Depletion Times}
\label{sec:mainsequence_deptime}
\begin{figure*}
\begin{center}
\includegraphics[width=0.48\textwidth,clip]{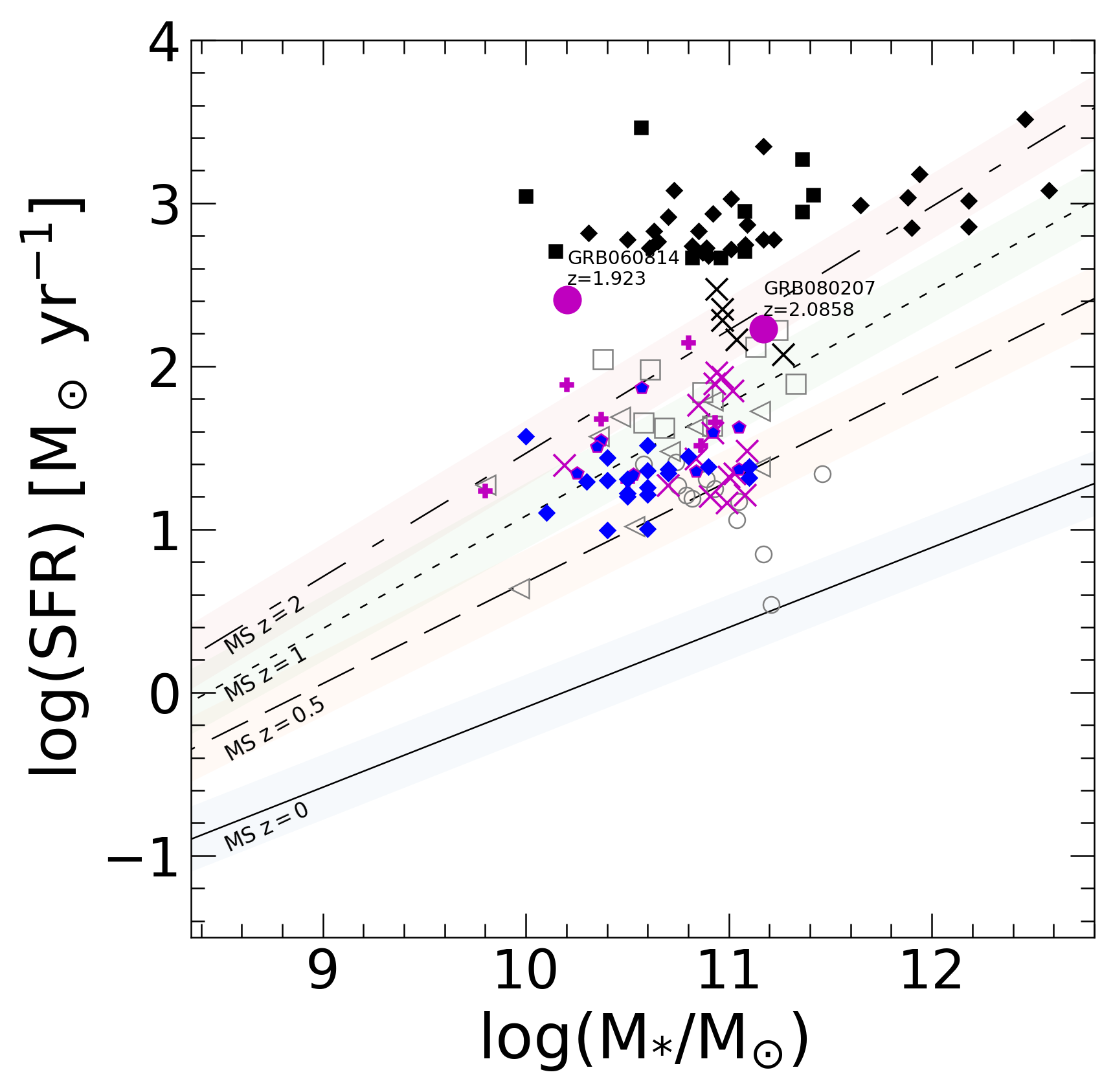} 
\includegraphics[width=0.48\textwidth,clip]{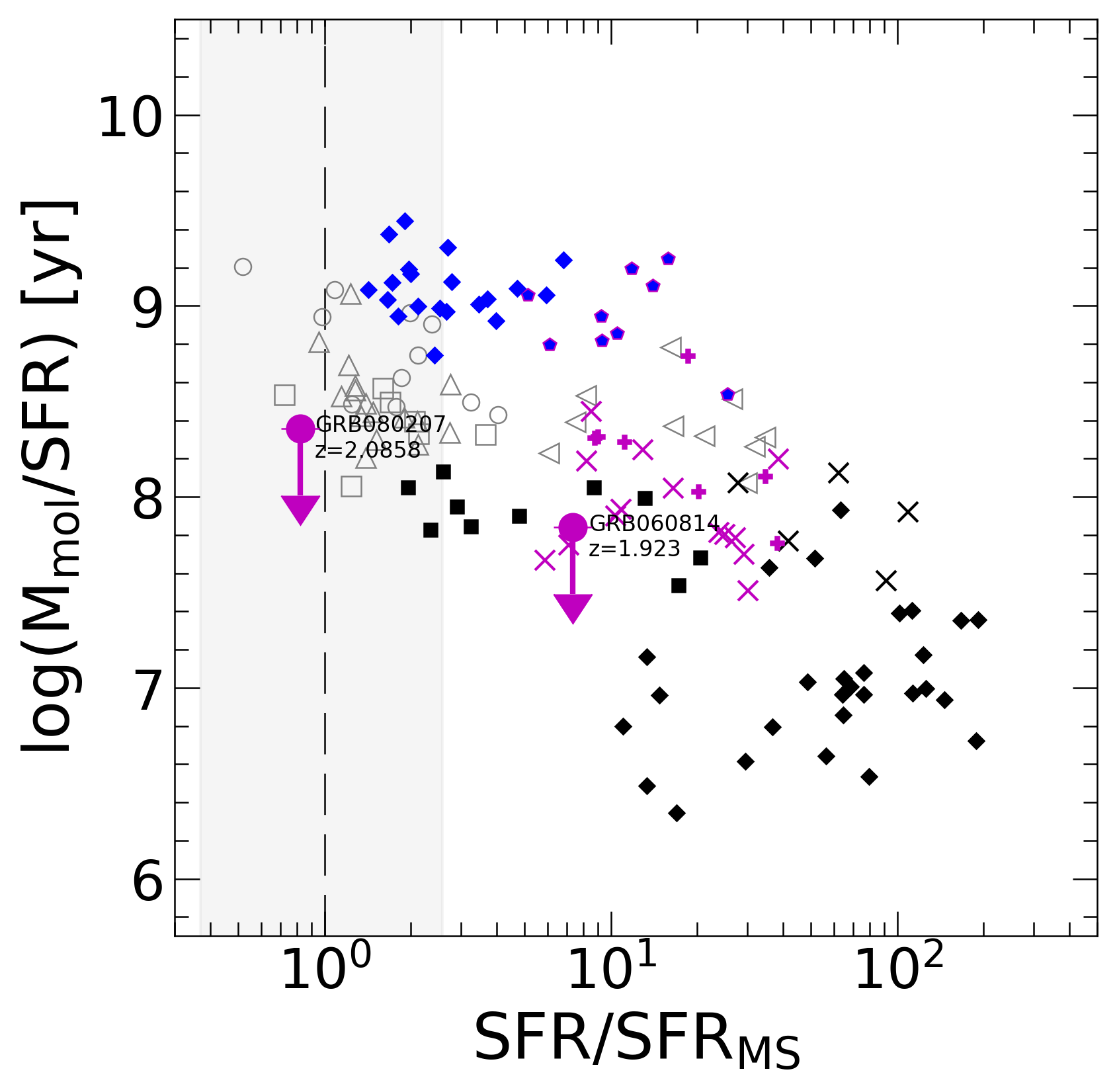}
\end{center}
\caption{Star-formation rate as a function of stellar mass (left panel), and depletion time as a function of distance from the MS SFR/SFR$_{\rm MS}$\ (right panel). All the symbols are the same as in Figure \ref{fig:Mmol_sfr}. The MS for the corresponding redshift has been estimated using results from \citet{Speagle2014ApJS..214...15S}. The shaded region represents 0.2 dex scatter. Black symbols correspond to galaxies that are in the starburst regime or will likely reach it thanks to their high gas masses.\label{fig:Mstar_SFR}}
\end{figure*}

To establish whether the position of GRB hosts in Figures \ref{fig:lpci_sfr} and \ref{fig:Mmol_sfr} is due to low \lpci\ or elevated SFRs, we investigate their location relative to the MS. Figure~\ref{fig:Mstar_SFR} shows the stellar mass as a function of SFR, and the MS for SFGs at different redshifts \citep{Speagle2014ApJS..214...15S}. 
While intermediate-$z$ LIRGs are found on their MS, the GRB hosts, (U)LIRGs, and transitioning SFG are found to lie above the MS for their redshifts. To better quantify this, the distance from the MS is estimated as the ratio of the observed SFR to the SFR of a galaxy on the MS with the same stellar mass and redshift (SFR/SFR$_{\rm MS}$). 

On the right panel of Figure \ref{fig:Mstar_SFR}, we show the molecular gas depletion timescale (\mmol/SFR using P04 method, or CO-based molecular gas masses from the literature when necessary), as a function of the distance to the MS. We find that  GRBs 060814 is about seven times above the MS, while GRB\,080207 lies on the corresponding MS within the scatter. The intermediate-$z$ isolated LIRGs lie close to the MS, while isolated transitioning galaxies are found one order of magnitude above it.
The transitioning merging galaxies and starburst merging ULIRGs are found at even greater distances (with an average offset of a factor 20 and 77, respectively).

We obtained relatively short gas depletion times of $<$64 and $<$238 Myr ({2$\sigma$ upper limits}) for the hosts of GRB\,060814, and 080207, respectively. These values are similar to those of high-$z$ starbursts for which we obtain a mean depletion time of 84 Myr and of transitioning mergers with an average of 205 Myr. The shortest depletion times (15 Myr on average) are found in intermediate-$z$ merging ULIRGs galaxies. As expected, the isolated LIRGs and isolated transitioning galaxies show longer depletion times of $\sim$1 Gyr on average.

In the right panel of Figure \ref{fig:Mstar_SFR} we can see that with the gradual departure from the MS, the depletion time 
decreases. 
\cite{Gao2004ApJ...606..271G} showed that the fraction of the dense gas is the primary predictor of the SFR, and that the relation between these quantities is  similar for different galaxy types. Here we can see that the merging process plays an important role in decreasing depletion time (or increasing SFE). Indeed, for the dense environment, it has been shown that the gas is converted quicker to stars, in particular in gas-rich mergers \citep{Genzel2010MNRAS.high_zSB_mergers}. {Perhaps this is to be expected given that in mergers there are much more turbulent molecular gas reservoirs, whose higher Mach numbers will place more gas at high densities ($\geq10^4\,{\rm cm}^{-3}$).}

\section{Discussion}
\label{sec:grb_hosts}
A common feature for both GRB hosts studied in this work is a sensitive \ci\ nondetection implying relatively low \ci\ emission. The disturbed multi-component morphology in the {\it Hubble Space Telescope} imaging of both galaxies suggests an ongoing merger process \citep{Svensson2012MNRAS.421...25S,Blanchard2016ApJ...817..144B,Chrimes2019MNRAS.486.3105C,Schneider2022A&A...666A..14S}. In what follow we put forward two non-exclusive scenarios to explain this feature. The first one is that the GRB hosts are caught at their transition from the main sequence toward the starburst phase \citep{Michiyama2021ApJS..257...28M,Hogan2022MNRAS.512.2371H,Cheng2018MNRAS.475..248C}. The second possibility is that they are \ci-dark galaxies, which may explain their low \ci/CO brightness temperature ratio \citep{Jiao2017ApJ...840L..18J,Michiyama2020ApJ...897L..19M}.

\subsection{Transition To Starburst}
{ Our GRB hosts exhibit lower molecular masses than the best fit to normal galaxies (Fig.~\ref{fig:Mmol_sfr}), similar to merging ULIRGs and high-$z$ starbursts. Moderate SFRs (as for the redshift of our GRB hosts) result in distances from their MS in between that measured for intermediate-$z$ isolated  LIRGs and that for intermediate- and high-$z$ starbursts. Estimated depletion times (\mmol/SFR) are similar to that of transitioning merger galaxies, while being shorter than intermediate-$z$ isolated LIRGs, and isolated transitioning galaxies, and longer (by about an order of magnitude) than the merging intermediate- and high-$z$ starburst galaxies (Fig.~\ref{fig:Mstar_SFR}, right panel). The moderate distances from the MS and short depletion times suggest that our GRB hosts are observed in their transition toward the starburst phase. Given the morphology, this increase in SFR is likely caused by mergers. The existence of GRB events indicates that this may be the beginning of such a transition, because the progenitors of GRBs are short-lived stars.}

{ We note that estimated depletion times assume no feedback effects (e.g.~energetic winds from massive stars, radiative pressure). These effects have been found to be a possible cause of the enlargement of the time over which starburst galaxies consume their available cold gas reservoirs. In this work we treat all the galaxies in the same manner, i.e. depletion times are estimated without such feedback effects \citep{Semenov_2017,Simone_Dep_Time2020A&A...635A.197D}, thus we consider this comparison as valid. 
Moreover, the depletion times are not used to draw any conclusions about the timescale of running out of gas.}

\subsection{[C I]-dark Galaxies}
{
The \ci\ nondetection may be caused by an underabundance of carbon, rather than low molecular hydrogen mass. This would result in a \ci-dark object.}
The \ci-dark galaxies are characterised by a low \ci/CO temperature brightness ratio, below 0.1 \citep{Michiyama2020ApJ...897L..19M,Michiyama2021ApJS..257...28M}, while regular ULIRGs have this ratio not lower than 0.2 \citep{Jiao2017ApJ...840L..18J}. These \ci-dark galaxies have high hydrogen densities between $10^5$ and $10^6$\,cm$^{-3}$, as shown using photodissociation region (PDR) models \citep{Valentino2020ApJ...890...24V,Michiyama2020ApJ...897L..19M,Michiyama2021ApJS..257...28M}.

In standard models of photodissociation regions, CO molecules are efficiently shielded from ultraviolet radiation at the dust extinction $A_V$ above a few mag, which could lead to low abundances of \ci\ \citep{Tielens1985ApJ...291..722T}. Recent models show, however, that the abundance ratio of \ci\ over CO  is linked more strongly with the cosmic rays ionization rate rather than the UV radiation field \citep{Bisbas2017ApJ...839...90B}. Nevertheless, \ci\ can be considered as a good tracer of \htwo\ gas in mergers, except in the highest-density medium where carbon is locked in CO. Such high densities are indeed commonly detected toward \ci-dark objects which have undergone a merger event \citep{Michiyama2020ApJ...897L..19M,Michiyama2021ApJS..257...28M}. 

{The proposed explanations of transitioning-to-starburst galaxies and \ci-dark galaxies may be related. The low \ci/CO line ratio in \ci-dark galaxies could be an effect of mergers yielding high average density H$_2$ gas reservoirs but with a yet-to-be-fully-ignited starburst, producing lower average cosmic ray energy densities. The combination of high $n_H$ and low cosmic ray energy density can then naturally produce \ci-dark galaxies, albeit only for short cosmic time intervals. In that regard, it would be interesting to examine whether \ci-dark galaxies of this type deviate from the far infrared-radio correlation, i.e., with lower synchrotron emission for a given far-infrared luminosity.}

The estimated { upper limit} \ci/CO temperature brightness ratio of $<0.1$ for the GRB 080207 hosts places it in the regime of \ci-dark galaxies. Very high gas density between $10^5$ and $10^6$\,cm$^{-3}$ have also been inferred for this galaxy using a \cii\ marginal detection \citep{Hashimoto2019MNRAS.488.5029H}, which is consistent with the proposed \ci-dark nature of the host. Both our GRB hosts have also low \ci\ luminosities for their SFRs.
In the case of the GRB\,060814 host, the \ci\ and CO were not detected,  so we can not confirm or rule out this interpretation for this galaxy. High gas densities were also claimed for other GRB hosts \citep{Christensen2008A&A...490...45C,Michalowski2014A&A...562A..70M,Michalowski2015A&A...582A..78M,Michalowski2016A&A...595A..72M,Arabsalmani2015MNRAS.454L..51A,Arabsalmani2019MNRAS.485.5411A,Arabsalmani2020ApJ...899..165A,Arabsalmani2022AJ....164...69A,deUgartePostigo2020A&A...633A..68D}. 

{ Thus, until additional observations (e.g.~of HCN to trace the dense gas phase) are available we conclude that both GRB hosts are candidates for \ci-dark galaxies that are transitioning toward the starburst phase with the merger event being the cause.}

\subsection{Ruled Out Mechanism: Post-starbursts}
In principle, our galaxies might be observed in the post-starburst phase, on the way down from the starburst regime. This would explain their SFRs and low gas content. However, this possibility is inconsistent with the presence of optical emission lines, indicating recent star formation. Furthermore, Balmer absorption features are not detected in the spectra of these GRB hosts \citep{Kruhler2015A&A...581A.125K}, unlike for post-starburst galaxies. Post-starburst galaxies have low SFRs and high metallicities, so while it is not impossible \citep[e.g.][]{Rossi2014A&A...572A..47R,Levan2023arXiv230312912L}, they are less likely to host a GRB, as opposed to an early starburst phase. 

\section{Conclusions}\label{sec:discussion_and_conclusions}
Our targets display spectra indicative of a young stellar population and have slightly increased SFRs relative to their MS  \citep{Perley2013ApJ...778..128P,Kruhler2015A&A...581A.125K}. Based on our \ci\ emission line measurements, we have inferred a molecular gas content similar to that of local ULIRGs but { lower than would be expected from the best fit to the normal star-forming galaxies (see Figs.~\ref{fig:lpci_sfr} and \ref{fig:Mmol_sfr}).}
{ We propose that these are high-$z$ merger systems caught at the transition from the main sequence toward the starburst phase. These are high redshift analogs to the intermediate-$z$ transitioning galaxies \citep{Cheng2018MNRAS.475..248C,Hogan2022MNRAS.512.2371H}. Furthermore, their low \ci/CO ratio point to a high-density environment observed in the collisional fronts of mergers \citep{Valentino2020ApJ...890...24V,Michiyama2020ApJ...897L..19M,Michiyama2021ApJS..257...28M}. Indeed, the merger signatures have been observed for our GRB hosts \citep{Svensson2012MNRAS.421...25S,Blanchard2016ApJ...817..144B,Chrimes2019MNRAS.486.3105C,Schneider2022A&A...666A..14S}. While such merger-driven starbursts play a lesser role in the overall star-formation rate density at $z=2$ \citep{Rodighiero2011ApJ...739L..40R}, these systems may play a crucial role in the quenching and morphological transformation.}

To test whether these galaxies indeed contain dense molecular star-forming clouds it would be essential to observe their emission of the high dipole-moment molecule HCN \citep{Gao2004ApJS..152...63G}. This, however, will be challenging because of the weakness of the HCN line. Additionally, the deep, high-resolution imaging of \ci, and CO (e.g., with the NOEMA interferometer) would be helpful to study the relative location of \ci\ and CO, which again is not an easy task considering that both GRB hosts are at $z\sim2$.

\section{Acknowledgments}
We thank the anonymous reviewer for all the comments that helped to improve our work. 
J.N., M.J.M., M.S.~and A.L.~acknowledge the support of 
the National Science Centre, Poland through the SONATA BIS grant 2018/30/E/ST9/00208. This research was funded in whole or in part by National Science Centre, Poland (grant number: 2021/41/N/ST9/02662). J.R.R. acknowledges support by grant PID2019-105552RB-C41 funded by MCIN/AEI/10.13039/501100011033. A.K. acknowledges support from the First TEAM grant of the Foundation for Polish Science No. POIR.04.04.00-00-5D21/18-00 and the Polish National Agency for Academic Exchange grant No. BPN/BEK/2021/1/00319/DEC/1. This article  has  been  supported  by  the Polish  National Agency  for  Academic Exchange  under  grant No. {PPI/APM/2018/1/00036/U/001}. A.R. acknowledges support from the INAF project Premiale Supporto Arizona \& Italia. J.H. was supported by a VILLUM FONDEN Investigator grant (project number 16599). Based on observations collected at the European Organisation for Astronomical Research in the Southern Hemisphere under ESO programmes 098.F-9300(A) and 098.D-024 (A). 
Based on observations with the Atacama Pathfinder EXperiment (APEX) telescope. APEX is a collaboration between the Max Planck Institute for Radio Astronomy, the European Southern Observatory, and the Onsala Space Observatory. Swedish observations on APEX are supported through Swedish Research Council grant No 2017-00648. 
This work is based on observations carried out under project number 172-16 with the IRAM 30m telescope. IRAM is supported by INSU/CNRS (France), MPG (Germany) and IGN (Spain). The research leading to these results has received funding from the European Union's Horizon 2020 research and innovation program under grant agreement No 730562 [RadioNet]. 
This research has made use of 
the GHostS database (\url{www.GRBhosts.org}), which is partly funded by Spitzer/NASA grant RSA Agreement No. 1287913; and the NASA's Astrophysics Data System Bibliographic Services.

\vspace{5mm}


\facility{APEX \citep{Gusten2006A&A...454L..13G}, IRAM-30m \citep{Carter2012A&A...538A..89C}}
\software{astropy \citep{2013A&A...558A..33A,2018AJ....156..123A}, GILDAS \citep{Pety2005sf2a.conf..721P} 
          }

\bibliography{bibl}{}
\bibliographystyle{aasjournal}


\listofchanges

\end{document}